\documentclass[journal,twoside,web]{ieeecolor}
\usepackage{jsen}
\usepackage{cite}
\usepackage{amsmath,amssymb,amsfonts}
\usepackage{algorithmic}
\usepackage{graphicx}
\usepackage{textcomp}
\usepackage{wrapfig}
\usepackage{subcaption}
\usepackage{multirow}

\def\BibTeX{{\rm B\kern-.05em{\sc i\kern-.025em b}\kern-.08em
    T\kern-.1667em\lower.7ex\hbox{E}\kern-.125emX}}

\begin{document}

\title{Detecting Drill Failure in the Small Short-sound Drill Dataset}
\author{Thanh Tran, Nhat Truong Pham, and Jan Lundgren
\thanks{This research was supported by the EU Regional Fund, the MiLo Project (No. 20201888), and the Acoustic sensor set for the AI monitoring systems (AISound ) project. \textit{(Thanh Tran and Nhat Truong Pham are co-first authors.)}}
\thanks{Thanh Tran and Jan Lundgren are with the Department of Electronics Design, Mid Sweden University, Sundsvall, Sweden (e-mails: thanh.tran@miun.se; jan.lundgren@miun.se).}
\thanks{Nhat Truong Pham is with the Institute for Computational Science, Ton Duc Thang University, Ho Chi Minh City, 72915, Vietnam (e-mail: phamnhattruong.st@tdtu.edu.vn).}}

\maketitle

\begin{abstract}
Monitoring the conditions of machines is vital in the manufacturing industry. Early detection of faulty components in machines for stopping and repairing the failed components can minimize the downtime of the machine. This article presents an approach to detect the failure occurring in drill machines based on drill sounds from Valmet AB. The drill dataset includes three classes: anomalous sounds, normal sounds, and  irrelevant sounds, which are also labeled as ``Broken", ``Normal", and ``Other", respectively. Detecting drill failure effectively remains a challenge due to the following reasons. The waveform of drill sound is complex and short for detection. Additionally, in realistic soundscapes, there are sounds and noise in the context at the same time. Moreover, the balanced dataset is small to apply state-of-the-art deep learning techniques. To overcome these aforementioned difficulties, we augmented sounds to increase the number of sounds in the dataset. We then proposed a convolutional neural network (CNN) combined with a long short-term memory (LSTM) to extract features from log-Mel spectrograms and learn global high-level feature representation for the classification of three classes. A leaky rectified linear unit (Leaky ReLU) was utilized as the activation function for our proposed CNN instead of the rectified linear unit (ReLU). Moreover, we deployed an attention mechanism at the frame level after the LSTM layer to learn long-term global feature representations. As a result, the proposed method reached an overall accuracy of 92.35\% for the drill failure detection system.
\end{abstract}



\section{Introduction}
\IEEEPARstart{D}{rill} fault detection systems are widely used in factories to prevent machine failure. The drilling machine is included 90 or 120 drill bits to drill thousand of small holes on the surface of the metal~\cite{9252126}. Broken drill bits can seriously damage a product because they lack holes on the metal surface. The maintenance technician stops the machine every 10 min to identify any broken drill bits and change them before re-initiating the drilling machine. Therefore, a fault detection system for the drilling machine is very crucial to minimize the downtime of the machine as well as the maintenance cost.

Many studies have been conducted on detecting and diagnosing drill failure in the past decade. Choi~\emph{et al.}~\cite{choi2008prediction} proposed a time domain and frequency domain feature extraction method named Characteristic Parameters of the Drill Failure (CPDF). In the second step, a multilayer perceptron (MLP) was used to predict drill failure based on the drill state index threshold. This would be lower the error rate. 
In order to solve this issue as well as improve the accuracy of diagnosis for drill failure, Skalle~\emph{et al.}~\cite{skalle2013detection} proposed a method based on symptom detection (e.g., Soft Fm, Cutting Accumulation, Local Dog Leg).
In~\cite{kumar2014detection}, Kumar~\emph{et al.} used vibration signals to detect and classify drill failures using three different classifier schemes: Artificial Neural Network (ANN), support vector machines (SVM), and Bayesian classifiers. Because vibration-based signals often contain noise, several techniques were required to remove noise and separate the sources to improve fault detection accuracy.

Researchers have used sound and vibration analysis in recent years to detect and classify faults~\cite{Henriquez2014}. As a result of the development of deep learning, CNNs (Convolutional Neural Networks) were used to automatically extract features to diagnose and classify faults on machines, especially drill machines. Due to the advantage of acoustic analysis over vibrations, Glowacz~\cite{glowacz2019fault} proposed an acoustic-based fault detection method for electric impact drills and coffee grinders. These acoustic features, including the root mean square (RMS) and a method selection of amplitude using a multi-expanded filter (MSAF-17-MULTIEXPANDED-FILTER-14), were used to classify fault status by the nearest neighbor classifier. Additionally, to detect the fault in electric impact drills, it is necessary to determine the fault of the gearbox device of the drill, since the gears are the main component of the power transmission. Jing~\emph{et al.}~\cite{jing2021fault} proposed a method for detecting electric impact drill failure by using logistic regression from time-varying loudness and acoustic signals.

Recently, a number of techniques have been investigated in the field of fault detection and machine condition monitoring. Hou~\emph{et al.}~\cite{hou2020fault} used wavelet packet energy to extract features from acoustic signals, then applied a feature selection method based on the Pearson correlation coefficient to select features. The selected features were used to classify the fault status with a neural network classifier. In addition to synchronous hydraulic motors, this approach can be applied to other rotating machines as well. In another approach, Wang~\emph{et al.}~\cite{wang2021bearing} proposed a multimodal method to detect bearing faults by fusing acoustic and vibration signals collected from the accelerometer and the microphone using the 1 dimensional CNN.

In recent years, deep learning has had a great deal of success in the detection and diagnosis of mechanical faults by using vibration and acoustics signals~\cite{polat2020fault},~\cite{gonzalez2020dcnn},~\cite{verstraete2017deep},~\cite{zhang2020deep},~\cite{chen2017vibration},~\cite{islam2018motor},~\cite{xueyi2020semi}.
Besides, recent studies have demonstrated that image representations of sound signals can be used to train the deep learning architecture for sound classification tasks. Researchers have proposed a lot of image representations for sounds, such as MFCC (Mel Frequency Cepstral Coefficients)~\cite{Zhang2017},~\cite{8635051}, spectrogram~\cite{Boddapati2017}, Mel spectrogram~\cite{Mushtaq2021}.
Additionally, many state-of-the-art deep learning models have been used for sound classification. Boddapati~\emph{et al.}~\cite {Boddapati2017} compared the classification accuracy of AlexNet and GoogleLNet on three different feature representations of sound (spectrogram, MFCC, and cross recurrence plot). A variant of conditional neural networks, called masked conditionaL neural network (MCLNN) has been proposed by Medhat~\emph{et al.}~\cite{Medhat2018} for classifying sounds. Researchers have used dilated CNNs with dilated filters and leaky ReLU activation functions~\cite{Zhang2017},~\cite{Chen2019}. The effect of modulating the dilation rate in dilated CNN on sound classification was compared in Chen~\emph{et al.}~\cite{Chen2019}. Recent studies have shown that recurrent neural networks (RNN) produce excellent results for variable-length sound sequences. Zhang~\emph{et al.}~\cite{Zhang2020} proposed a CNN architecture to learn spectro-temporal features and a bidirectional gated recurrent unit (Bi-GRU) with a frame-level attention mechanism for sound classification. Wang~\emph{et al.}~\cite{Wang2019} proposed a CNN architecture with a parallel temporal-spectral attention mechanism to capture certain frames where sound events occur and pay attention to varying frequency bands.

Our article proposed an approach to detect drill machine failures based on drill sounds from Valmet AB. Most drilling fault detection studies, however, used a large, balanced dataset. In fact, the anomalous sounds that were captured while drill machines were damaged are only a small fraction of the overall dataset. Deep learning models might not do well in the prediction of minority classes due to imbalanced datasets in real-world applications. Also, the extracted features from raw sound signals are insufficient for classification since the sample durations for sounds in our dataset are around 20.83 ms and 41.67 ms. This makes it more challenging to compare our results to those of previous research in the field of drilling sounds classification. As a result, an end-to-end deep learning system faces many challenges when it comes to detecting drill faults. To overcome these difficulties, we re-sampled the dataset (undersampling) as described in~\cite{9252126}. Then we applied data augmentation to generate more samples of the dataset. The augmentation methods were shifting the sound by 5 ms and increasing the volume by 2. These sounds in the augmented dataset were converted into log-Mel spectrograms. In addition, we proposed a CNN combined with an attention-based LSTM for classifying drill sounds. Feature maps were extracted from the log-Mel spectrograms using CNN, and then an LSTM layer was used to learn high-level global feature representation from extracted features. We used Leaky ReLU in CNN instead of using ReLU to alleviate the potential problem that CNN stops learning when the ReLU has a value of less than zero. Leaky ReLU helps CNN continue learning when input values are negative. To focus on the important parts of drill sounds and discard the unnecessary parts, an attention layer was added after the LSTM.

The rest of the article is organized as follows. Our proposed method is described in Section II. Section III presents experiment results and discussion. Finally, section IV is the conclusion.

\section{Proposed Methodology}
Our proposed architecture is described as shown in Fig.~1. Firstly, original sounds were applied audio augmentation methods to increase the number of samples in the dataset. In the next step, we proposed a small CNN architecture that includes five layers to generate features from the Mel spectrograms of sounds. Finally, these features were used as the input of the LSTM with the attention mechanism to learn high-level feature representation. The details of the layers in our proposed model are described in Table~\ref{tablemain}, where $nC$ is the number of classes.
\begin{figure*}
\centering
  \includegraphics[width=0.95\linewidth]{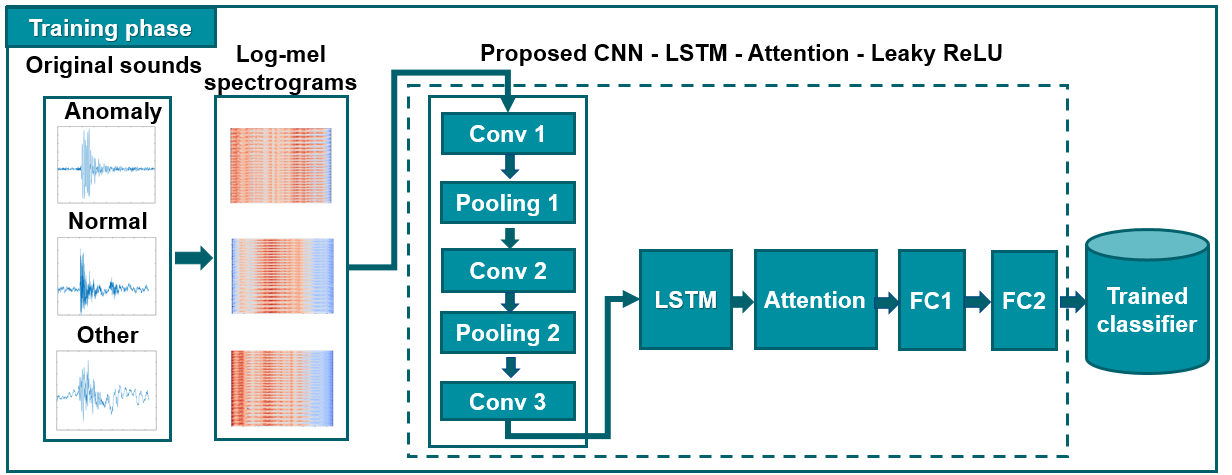}
  \caption{Our proposed methodology.}
  \label{fig:boat1}
\end{figure*}

\begin{table}[ht]
    \centering
    \caption{The layers of the proposed model.}
    \label{tablemain}
    \begin{tabular}{c|c|c}
        \hline
        \textbf{Layer} & \textbf{Kernel/Size} & \textbf{Output shape} \\ \hline
        Input          & ---             & 100\(\times\)96\(\times\)1       \\ \hline
        Conv 1          & 3\(\times\)3            & 100\(\times\)96\(\times\)128     \\ \hline
        Max\_pooling 1 & 2\(\times\)4             & 50\(\times\)24\(\times\)128      \\ \hline
        Conv 2          & 3\(\times\)3             & 50\(\times\)24\(\times\)128      \\ \hline
        Max\_pooling 2 & 2\(\times\)4             & 25\(\times\)6\(\times\)128       \\ \hline
        Conv 3          & 3\(\times\)3             & 25\(\times\)6\(\times\)256       \\ \hline
        LSTM        & 256             & ---       \\ \hline
        Attention       & ---             & ---            \\ \hline
        FC 1           & ---             & 64             \\ \hline
        FC 2           & ---             & nC             \\ \hline
    \end{tabular}
\end{table}

\subsection{Data Augmentation}
This study is a continuation of the study presented in \cite{9252126}. Therefore, we used the same dataset. Valmet AB drills small holes in metal plates with multiple machines. There are two type of drilling machines in a factory that are 90 and 120 bits. Fig.~2 shows a healthy drill bit and a broken drill bit \cite{9252126}.
\begin{figure}
\centering
\includegraphics[width=0.95\linewidth]{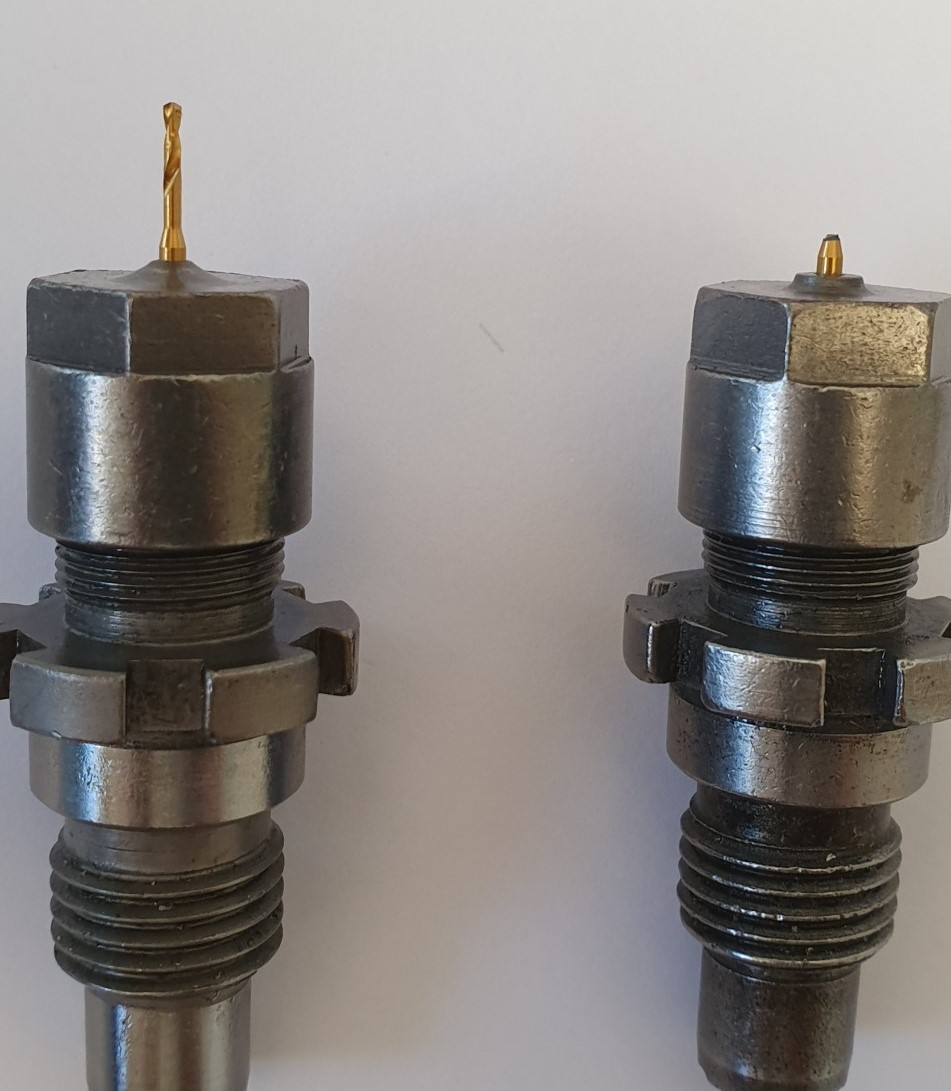}
\caption{A healthy drill bit (on the left side) and a broken drill bit (on the right side) \cite{9252126}.}
\end{figure}
In this dataset, sound from a drill machine in Sundsvall, Sweden was recorded with four AudioBox iTwo Studio microphones. For capturing drill sounds, 96 kHz was used as the sampling rate. 
The dataset has two parts. There are 833 files in one part. In this part, the length of each sound is 20.83 ms. The other part contains 40,417 files. In this part, the length of each sound is 41.67 ms.
On average, broken drill bits account for just 0.16\% of the total number of drill sounds recorded (67 anomalous sounds out of 41,250 total drill sounds). Therefore, broken drills are hard to detect. To proceed further, it is therefore important to create a balanced dataset. In the majority classes (normal sounds and unrelated sound classes), we selected randomly the same amount of sounds as in the minority classes (anomalous sounds). We have 201 sounds in the balanced dataset after re-sampling. There is a limited amount of data to work within the balanced dataset.

Although we tuned hyper-parameters of our model to adapt to the drill fault detection task, however, the lack of sound when the drills were broken is still a big challenge. 
To tackle this challenge, we applied data augmentation to the original sounds. Thus, the number of sounds in the dataset increased.
Besides, data augmentation helps improve the generalization capability of our proposed model. 
There are many methods of audio augmentation such as time-stretching, pitch-shifting, volume control, noise addition, etc. It is not appropriate to apply some augmentation methods to the sounds in our dataset since they are very short at only 20.83 ms or 41.67 ms. 
Experimentation revealed that only time-shifting and volume control data augmentation methods are effective for our data set.

In this article, we applied time-shifting and volume control to generate syntactical sounds. We did not add noise to the sound as an augmentation method because the sound in our dataset is very short. Noise makes it difficult to classify sounds. Matlab provides a simple function, audioDataAugmenter to augment the sound.

\subsubsection{Time-shifting} \text{}

A time-shifting is the process of shifting the sound backward or forward at random. We shifted the starting point of the sound by 5 ms to the right, then padded it back to its original length. Fig.~3a shows the time representation of the original fault sound and augmented sound using time-shifting. 

\begin{figure*}[h!]
  \centering
  \begin{subfigure}[b]{0.45\linewidth}
    \includegraphics[width=\linewidth]{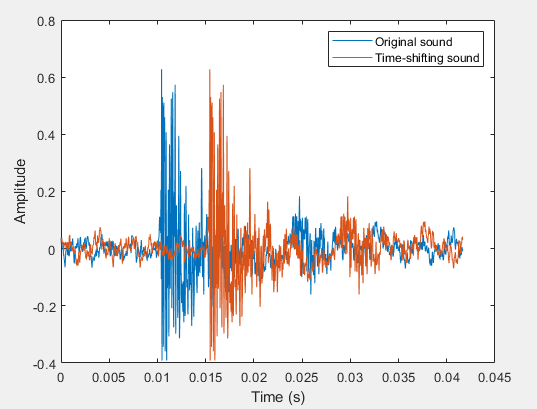}
     \caption{Time representation of the original fault sound and augmented sound using time-shifting.}
  \end{subfigure}
  \begin{subfigure}[b]{0.45\linewidth}
    \includegraphics[width=\linewidth]{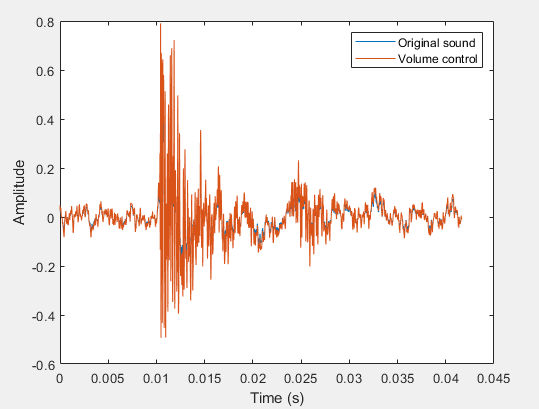}
    \caption{Time representation of the original fault sound and the augmented sound using volume control.}
  \end{subfigure}
  \caption{The time representation of the original fault sound and the augmented sound using time-shifting and volume control.}
  \label{fig:figure1}
\end{figure*}

\begin{figure*}[h!]
  \centering
  \begin{subfigure}[b]{0.3\linewidth}
    \includegraphics[width=\linewidth]{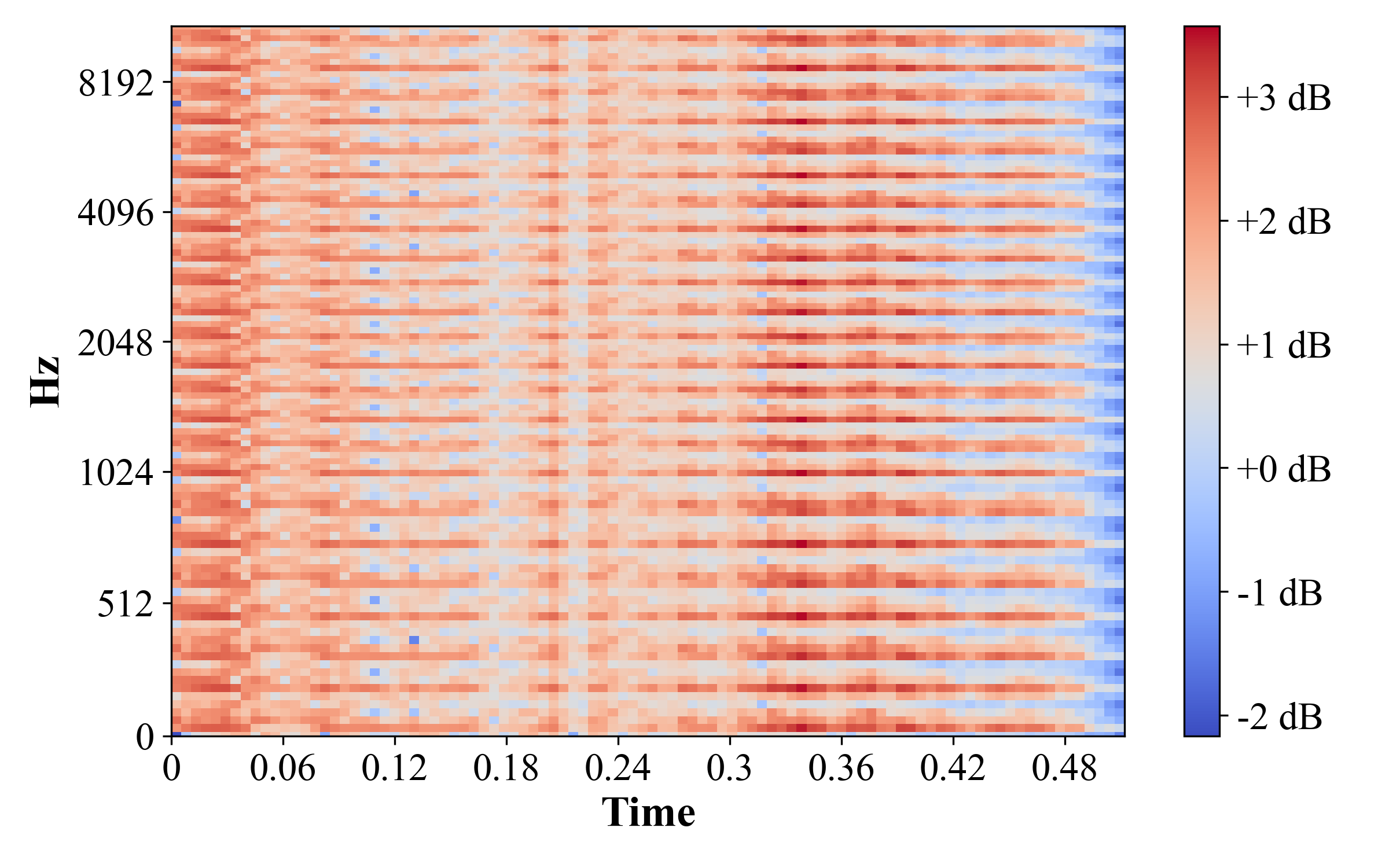}
     \caption{An anomalous sound (a broken drill).}
  \end{subfigure}
  \begin{subfigure}[b]{0.3\linewidth}
    \includegraphics[width=\linewidth]{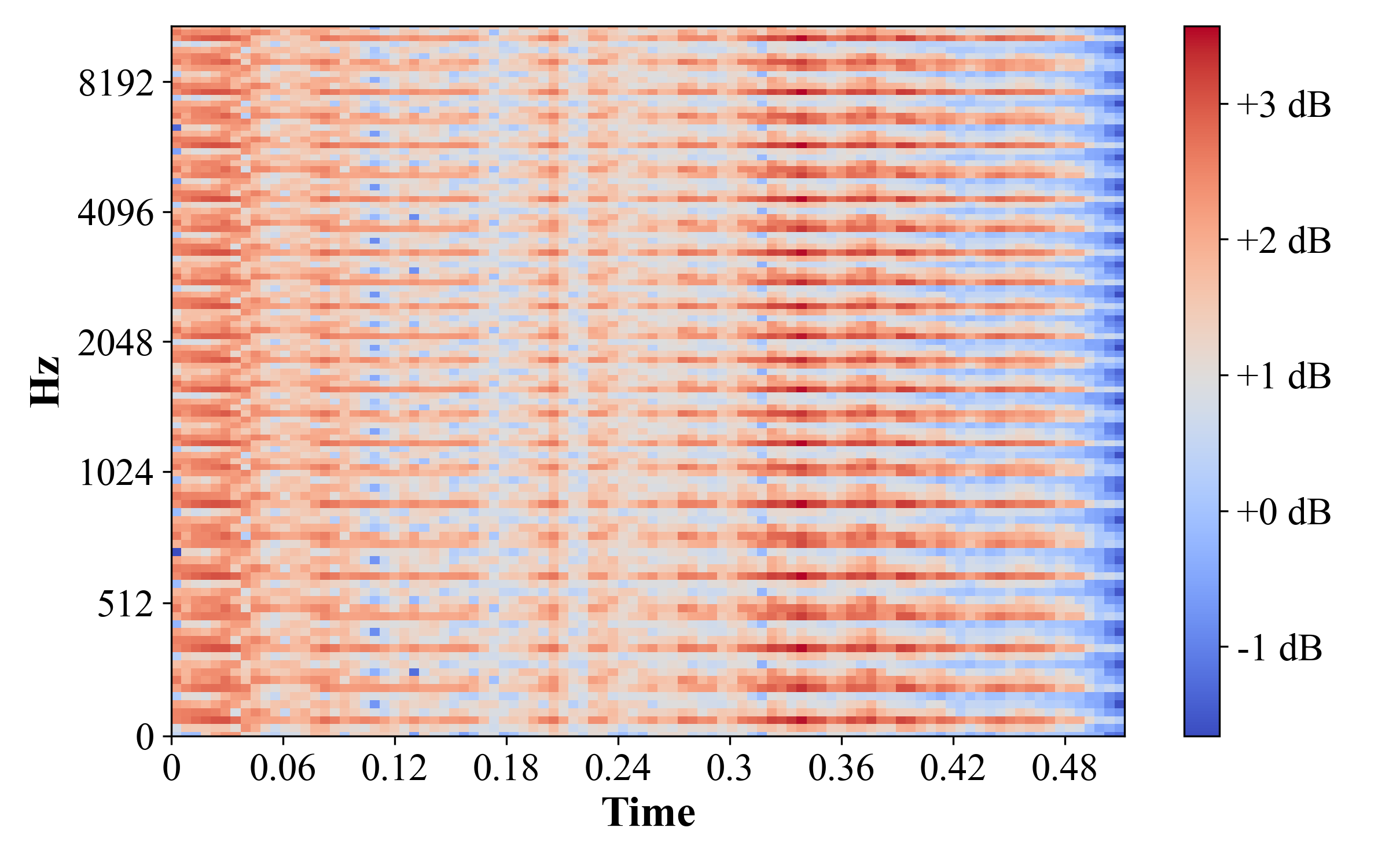}
    \caption{A time-shifted sound.}
  \end{subfigure}
  \begin{subfigure}[b]{0.3\linewidth}
    \includegraphics[width=\linewidth]{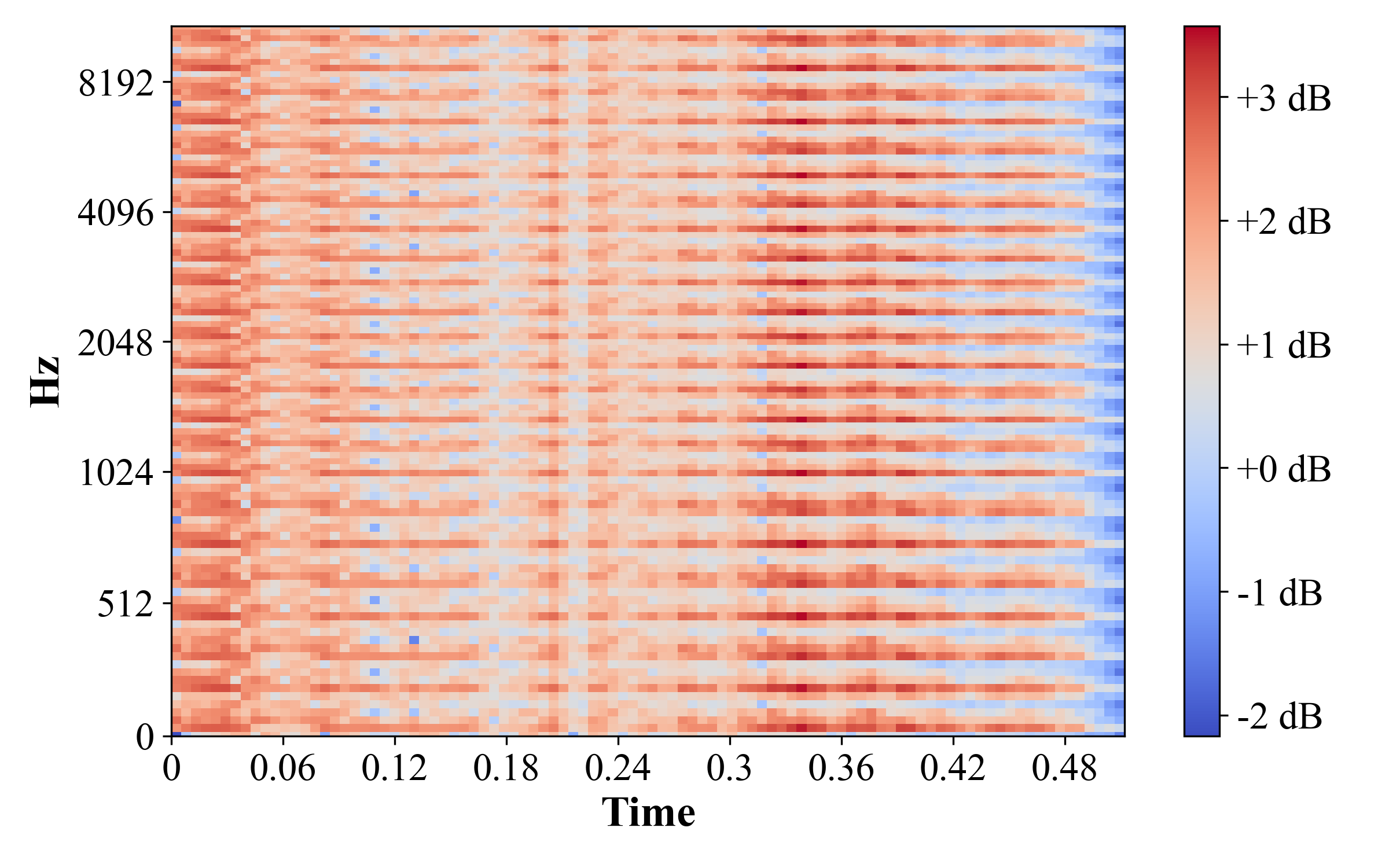}
    \caption{A volume gained sound.}
  \end{subfigure}
  \caption{Log-Mel spectrograms of an original anomalous sound, the augmented sound using time-shifting, the augmented sound using volume control.}
  \label{fig:figure1}
\end{figure*}

\subsubsection{Volume control} \text{} 

We increased the volume by multiplying the audio by a random amplitude factor. The volume gain was set as 2 dB. Using this technique, we can gain some in-variance concerning the audio input gain. In Fig. 3b, we show the time representation of the original fault sound and augmented sound using the volume control. 

\subsection{Convert Sounds into Log-Mel Spectrograms}
Recent advances in the field of image classification using CNN for multiple classes with high accuracy motivated us to investigate the ability of image representation of sounds to detect drill failures. In this paper, drill sounds were converted into log-Mel spectrograms to feed into our proposed CNN. The log-Mel spectrogram was generated as follows. Given a raw drill sound, we computed the Mel spectrogram using short-time Fourier transform (STFT) with Hamming windows of 100 ms and the hop length of 50 ms, the length of FFT was 2,048, the sampling rate was 96 kHz, and the number of Mel-filter bank was 96. Since the authors in \cite{choi2018comparison} found that the log-scaled Mel spectrogram improves the classification accuracy compared to the Mel spectrogram. Therefore, we took the logarithm of the Mel spectrogram as the input of our proposed CNN architecture. Fig.~4 shows log-Mel spectrograms of an original anomalous sound and its augmented sound using volume control and time-shifting.


\subsection{Extract Features using CNN with Leaky ReLU}
We proposed a CNN architecture only used for extracting features from log-Mel spectrograms. As a result, we used the third convolutional layer for extracting features instead of adding a dense layer at the end. Additionally, instead of using ReLU as the activation function, we used the Leaky ReLU. The experiment results show that using Leaky ReLU can improve the classification accuracy of our dataset. To learn global high-level feature representation, the extracted features were fed into LSTM with an attention mechanism.

Our proposed CNN architecture consisted of three convolutional layers and two max-pooling layers, and six batch normalization layers with the Leaky ReLU activation functions, as shown in Fig. 5. Log-Mel spectrograms were fed into our proposed CNN to extract high-level features for the classification task. Firstly, we utilized three convolutional layers with 3\(\times\)3 filter kernel sizes. Three convolutional layers have 128, 128, and 256 feature maps, respectively. Secondly, we added a max-pooling layer with 2\(\times\)4 filter kernel sizes after the first two convolutional layers. We also added a pair of batch normalization (BN) layers with Leaky ReLU before and after the convolutional layers to normalize the features and reduce over-fitting.

\begin{figure*}
\centering
  \includegraphics[width=0.75\linewidth,height=7cm]{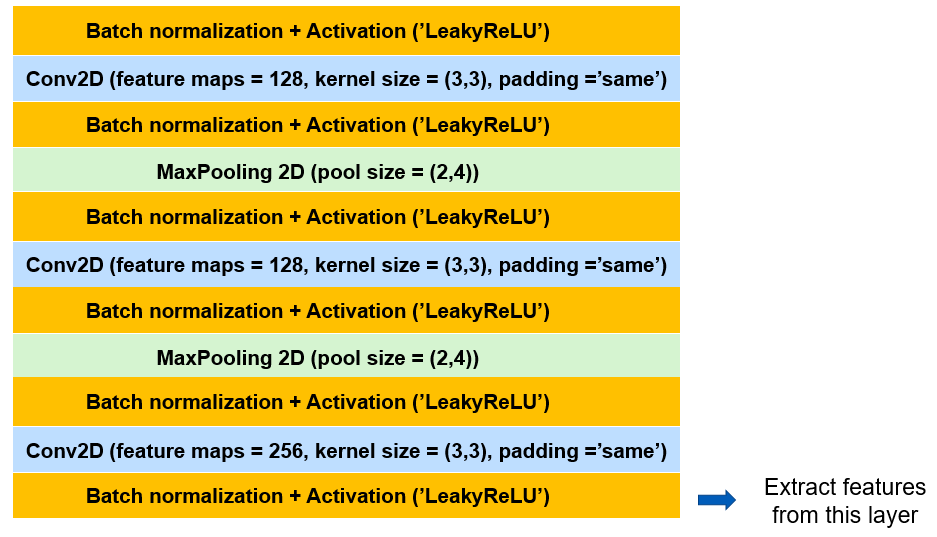}
  \caption{Our proposed CNN with Leaky ReLU.}
  \label{fig:boat1}
\end{figure*}

The equation for ReLU is~\( f(x) = max(0,x)\). When the input of the layer is negative, the ReLU is equal to zero. Consequently, gradient descents reach the value of zero and cannot converge to the local minimum. For Leaky ReLU, there is always a small slope to allow the weight update of the accumulated gradient. Therefore, although ReLU can compute faster, we used Leaky ReLU instead of the ReLU so that the layers did not stop learning when the slope of the ReLU is zero. The Leaky ReLU activation function \cite{Maas2013} is described by the equation 1:
\begin{equation}
    f(x)=
    \begin{cases}
      x, & \text{if}\ x>0 \\
      \alpha x, & \text{otherwise},
    \end{cases}
\end{equation}
where $\alpha$ was set to 0.3 in this research.


\subsection{Global Feature Learning using LSTM and Attention Mechanism}
In this article, we utilized LSTM \cite{hochreiter1997long} to learn sequential feature maps that are extracted from our proposed CNN. Each LSTM unit can be updated as in equations 2 -- 7:
\begin{equation}
    f_{t} = \sigma\big(W_{f}X_{t} + U_{f}h_{t-1} + b_{f}\big),
\end{equation}
\begin{equation}
    i_{t} = \sigma\big(W_{i}X_{t} + U_{i}h_{t-1} + b_{i}\big),
\end{equation}
\begin{equation}
    o_{t} = \sigma\big(W_{o}X_{t} + U_{o}h_{t-1} + b_{o}\big),
\end{equation}
\begin{equation}
    \tilde{c}_{t} = \tau\big(W_{c}X_{t} + U_{c}h_{t-1} + b_{c}\big),
\end{equation}
\begin{equation}
    c_{t} = f_{t}\odot\tilde{c}_{t-1} + i_{t} \odot\tilde{c}_{t},
\end{equation}
\begin{equation}
    h_{t} = o_{t} \odot \tau (c_{t}),
\end{equation}
where all variables are defined as follows:
\begin{itemize}
    \item $X_{t}$ is the mini-batch input;
    \item $i_{t}$ is the input gate;
    \item $f_{t}$ is the forget gate;
    \item $o_{t}$ is the output gate;
    \item $\tilde{c}_{t}$ is the input cell;
    \item $c_{t}$ is the cell state;
    \item $h_{t}$ is the hidden state;
    \item $\sigma$ is the \emph{sigmoid} function;
    \item $\tau$ is the \emph{tanh} function;
    \item $W$,~$U$ are the weight matrices;
    \item $b$ is the bias parameter;
    \item $t$ is the time step.
\end{itemize}

Due to different frame-level features contribute unequally to classifying event sound classes. Hence, we added an attention mechanism layer \cite{vaswani2017attention} after LSTM to specific points in a sequence when computing its output. 
Additionally, during the transition from the normal state of the drill bit to the broken state, the pitch of the audio changes. Therefore, the features extracted from the log-Mel spectrogram right at the moment the drill bit cracks will have an abnormality. The purpose of the attention layer is to focus on that anomaly.
The attention mechanism has been widely used in the sequence-to-sequence model. For the LSTM, the output of attention $att$ can be defined as below:
\begin{equation}
    att = \sum_{t=1}^{T}\alpha_{t}h_{t},
\end{equation}
where $h_{t}$ denotes the $t_{th}$ hidden output from the LSTM at time step $t$, $T$ represents the total number of time steps in the input sequence, and the $\alpha_{t}$ is the attention weight can be computed as follows:
\begin{equation}
    \alpha_{t} = \frac{\exp{\big(W \cdot h_{t}\big)}}{\sum_{k=1}^{T} \exp{\big(W \cdot h_{t}\big)}}.
\end{equation}

\section{Experimental Results}
\subsection{Experimental Setup}
\subsubsection{Datasets} \text{}

After applying time-shifting and volume control augmentation methods to 201 original sounds from three categories (anomaly, normal, and irrelevant sounds), we obtained 603 sounds. To train our end-to-end model, we converted these sounds in the augmented dataset into log-Mel spectrograms. We used around  70\%  of the dataset  (420 log-Mel spectrograms)  for training and  30\%  (183 log-Mel spectrograms) for testing, in which when training the model, 420 was divided by the ratio of 70/30 again for training and validation sets.


\subsubsection{Hyper-parameters and training setup} \text{}

The proposed deep learning model was trained on Intel CORE i5 8th Gen with NVIDIA Graphics Card 1050Ti. We used Keras library~\cite{chollet2015keras} with TensorFlow toolkit~\cite{tensorflow2015-whitepaper} that are popular deep learning frameworks to implement and deploy our proposed deep learning model. Additionally, the Librosa library~\cite{brian_mcfee_2021_4792298} was used to generate log-Mel spectrograms from original drill sounds.

For hyper-parameters optimization, we used the Adam optimizer~\cite{AdamKingma} with a learning rate of 0.001, a batch size of~4, a momentum of 0.9, and 100 epochs. During training, categorical cross-entropy was used as the loss function $L_{f}$ to update the network weights. It is defined as follows:

\begin{equation}
    L_{f} = - \sum_{n=1}^{nC}y_{n} \log\big(\hat{y}_{n}\big),
\end{equation}
where $nC$ is the number of classes, $y_{n}$ is the ground truth, and $\hat{y}_{n}$ is the predicted class probabilities for the $n_{th}$ element of model predictions. Furthermore, to avoid over-fitting and to improve the generalized model, we applied early stopping to train the network with the patience of 5.

\subsection{Results}

As shown in Table III, the overall accuracy of our proposed method, CNN using the Leaky ReLU activation function combined with attention-based LSTM (CNN - LSTM - Attention - Leaky ReLU), was 92.35\%. The confusion matrix for our proposed method is in Fig. 6.  Table II shows the F1-score, precision, and recall for each class in the augmented dataset. 

Our previous work in~\cite{9252126} used the Continuous wavelet transform (CWT) and the Mel spectrogram as the input of machine learning classifiers (linear discriminant, SVM, and KNN). In contrast, in this paper, we proposed an end-to-end model that utilized log-Mel spectrograms as inputs and produced an accuracy of 92.35\%, whereas the previous work produced an accuracy of 80.25\%. 
\begin{figure}
\centering
  \includegraphics[width=0.95\linewidth]{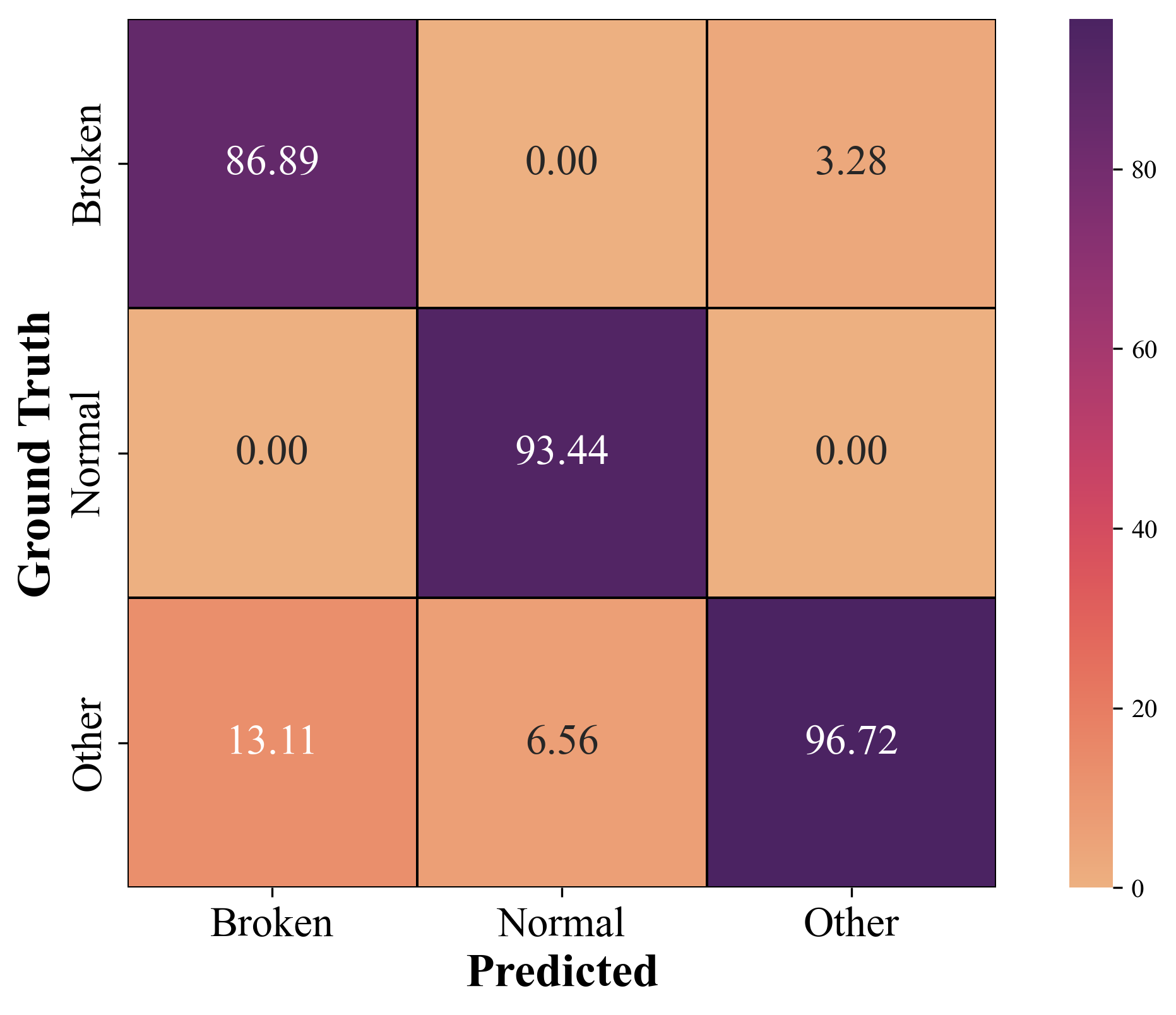}
  \caption{The confusion matrix for the proposed model (CNN - LSTM  - Attention - Leaky ReLU) on the augmented dataset.}
  \label{fig:result3}
\end{figure}

\begin{table}[ht]
\centering
\caption{The classification results of CNN with Leaky ReLU in conjunction with attention-based LSTM}.
\label{tablere3}

\begin{tabular}{l|c|c|c|c}
\hline
\multicolumn{1}{c|}{}                                             & \textbf{Precision }  & \textbf{Recall }  & \textbf{F1-score }  & \textbf{No.}   \\ 
\hline
Anomaly sound (``Broken")        & 0.96                 & 0.87              & 0.91                & 61                       \\ 
\hline
Normal sound (``Normal")                                                         & 1.00                 & 0.93              & 0.97                & 61                       \\ 
\hline
Irrelevant sound (``Other")                                                              & 0.83                 & 0.97              & 0.89                & 61                       \\ 
\hline
\textbf{Accuracy }                                                &                     &                  & \textbf{0.92 }     & \textbf{183}     \\ 
\hline
\textbf{Macro avg }                                               & \textbf{0.92 }      & \textbf{0.93 }   & \textbf{0.92 }     & \textbf{183}     \\ 
\hline
\multicolumn{1}{l|}{\textbf{Weighted avg }}                       & \textbf{0.92 }      & \textbf{0.93 }   & \textbf{0.92 }     & \textbf{183}     \\
\hline
\end{tabular}
\end{table}

\subsection{Discussion}
This part aims to demonstrate that data augmentation, the Leaky ReLU and the attention mechanism can affect the overall accuracy of the proposed method when combined with CNN and LSTM. The mean accuracy of all the experiments is shown in Table III for comparison. Using CNN with Leaky ReLU activation function in conjunction with attention-based LSTM achieves the highest accuracy of 92.35\%. The following experiments were conducted to validate the effectiveness of our proposed method:

\begin{table}[]
\centering
\caption{The comparison of different models.}
\label{tablecompa}
\begin{tabular}{l|c}
\hline
\textbf{Model}                  & \textbf{Mean Accuracy (\%)} \\ 
\hline
CNN - Leaky ReLU                & 69.95                       \\ \hline
CNN - LSTM - Leaky ReLU         & 86.89                       \\ \hline
\textbf{CNN - LSTM - Attention - Leaky ReLU}   & \textbf{92.35}                       \\ \hline
CNN - LSTM - Attention - ReLU         & 90.16                       \\ \hline
\end{tabular}
\end{table}
\subsubsection{CNN - Leaky ReLU} \text{} 

In CNN architecture, we run experiments with the Leaky ReLU activation function, which has layers similar to Fig.~5. The experiment parameters were identical to the CNN architecture in our proposed method. However, we used two fully connected layers at the end of CNN for classification. According to Table V, the overall accuracy for this method was only 69.95\%, which is lower than the accuracy of our proposed method (92.35\%). Fig.~7 shows the confusion matrix for this method.

\begin{figure}
\centering
  \includegraphics[width=0.95\linewidth]{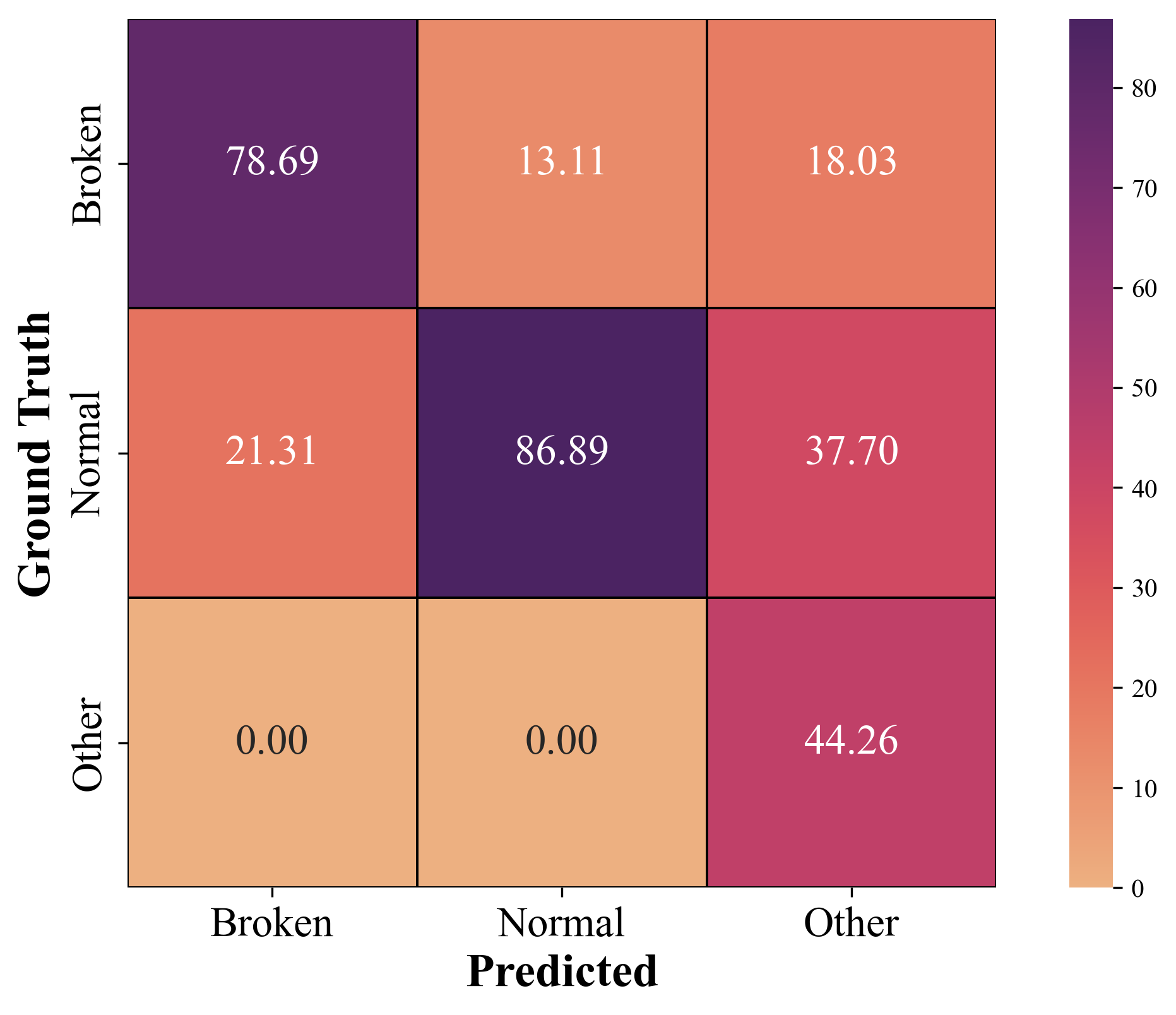}
  \caption{CNN - Leaky ReLU.}
  \label{fig:result1}
\end{figure}

\subsubsection{CNN - LSTM - Leaky ReLU} \text{} 

This part experimented with CNN uses the Leaky ReLU activation function in conjunction with LSTM.  This experiment tests whether incorporating an attention layer into the model is effective. In this method, accuracy achieved 86.89\%, which is less than what we propose in section IIIB (accuracy of 92.35\%). It is clear that the accuracy of the model was improved by including the attention layer. In theory, with the attention layer, the LSTM is supposed to invest more computing power of that small but important part of the input, so the network enhances these parts and fades out the rest. The confusion matrix for this method is shown in Fig. 8. 

\begin{figure}
\centering
  \includegraphics[width=0.95\linewidth]{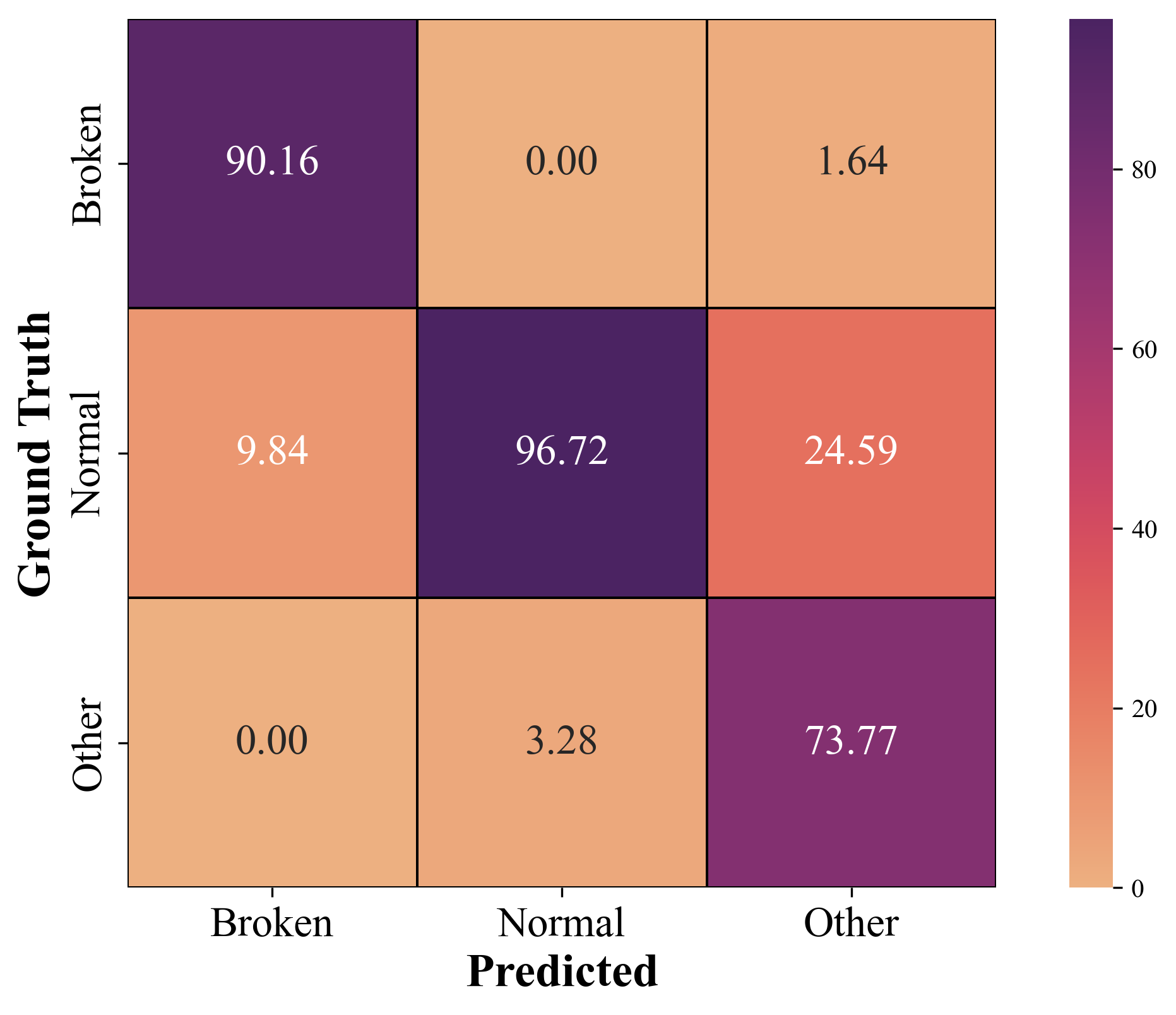}
  \caption{CNN - LSTM - Leaky ReLU.}
  \label{fig:result2}
\end{figure}

\subsubsection{CNN  -  LSTM  -  Attention  -  ReLU} \text{}

This part experimented with CNN architecture uses the ReLU activation function in conjunction with attention-based LSTM. The confusion matrix for this method is shown in Fig.~9. 
In this experiment, we run the model with ReLU activation to prove it is less effective than Leaky ReLU activation on our dataset.
When using the ReLU activation function, the accuracy was 90.16\%, while using Leaky ReLU, the accuracy was higher (92.35\%). As Leaky ReLU has a slope of 0.3 instead of 0, CNN can train faster and avoid the ``dying ReLU" problem on our dataset.

\begin{figure}
\centering
  \includegraphics[width=0.95\linewidth]{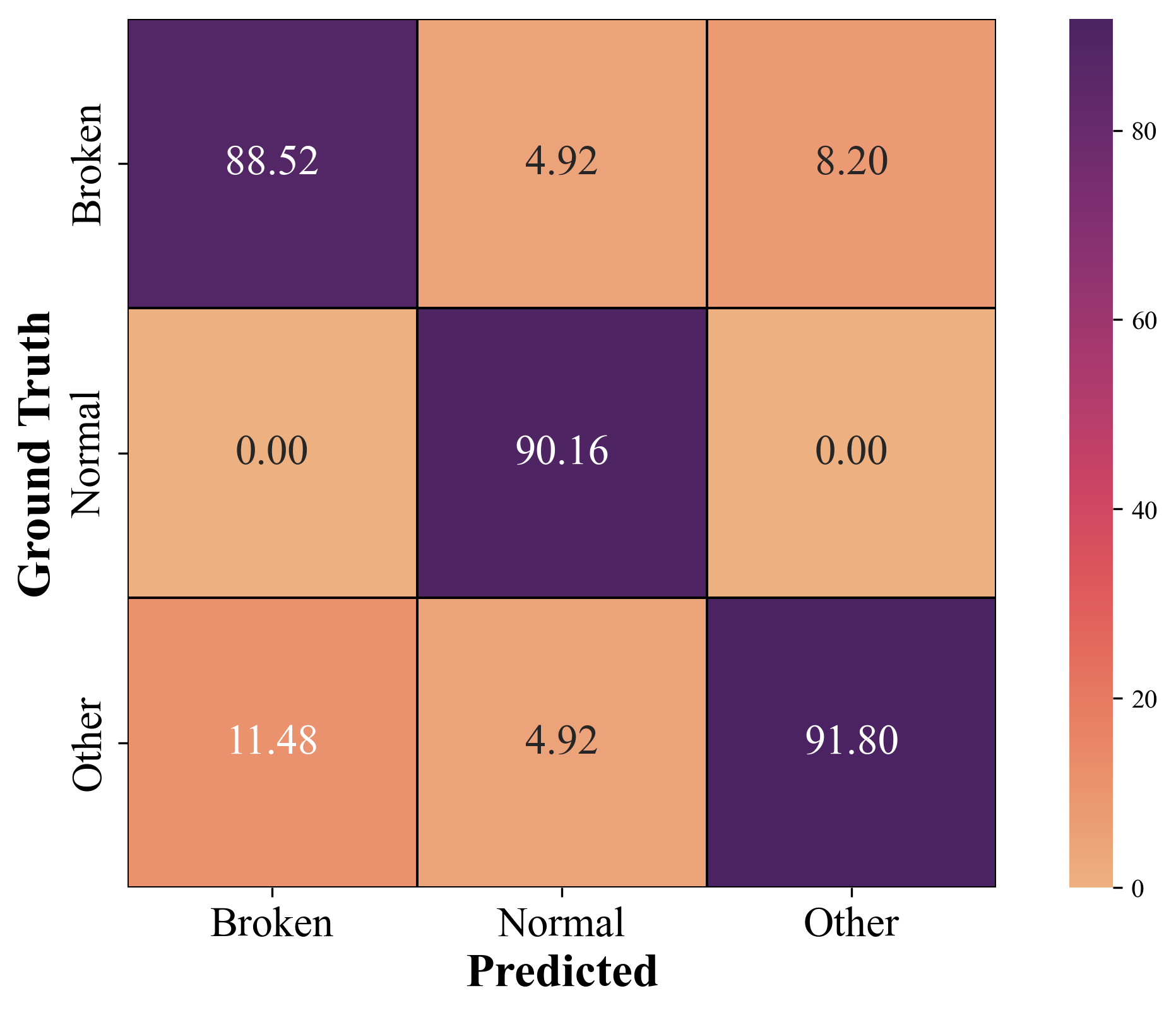}
  \caption{CNN - LSTM - Attention - ReLU.}
  \label{fig:result4}
\end{figure}

\subsubsection{The comparison between using the proposed method on the original dataset and the augmented dataset} \text{}

To test the efficiency of the data augmentation process, the proposed model in section II was run on both the original and augmented datasets. The results are shown in Table IV, the accuracy on the augmented dataset reached 92.35\% whereas the accuracy on the original dataset only reached 84.13\%. The accuracy of our proposed method on the augmented dataset (603 sounds) is clearly higher than on the original dataset (201 sounds).
The confusion matrix for our proposed method on the original dataset is shown in Fig. 10.

\begin{figure}
\centering
  \includegraphics[width=0.95\linewidth]{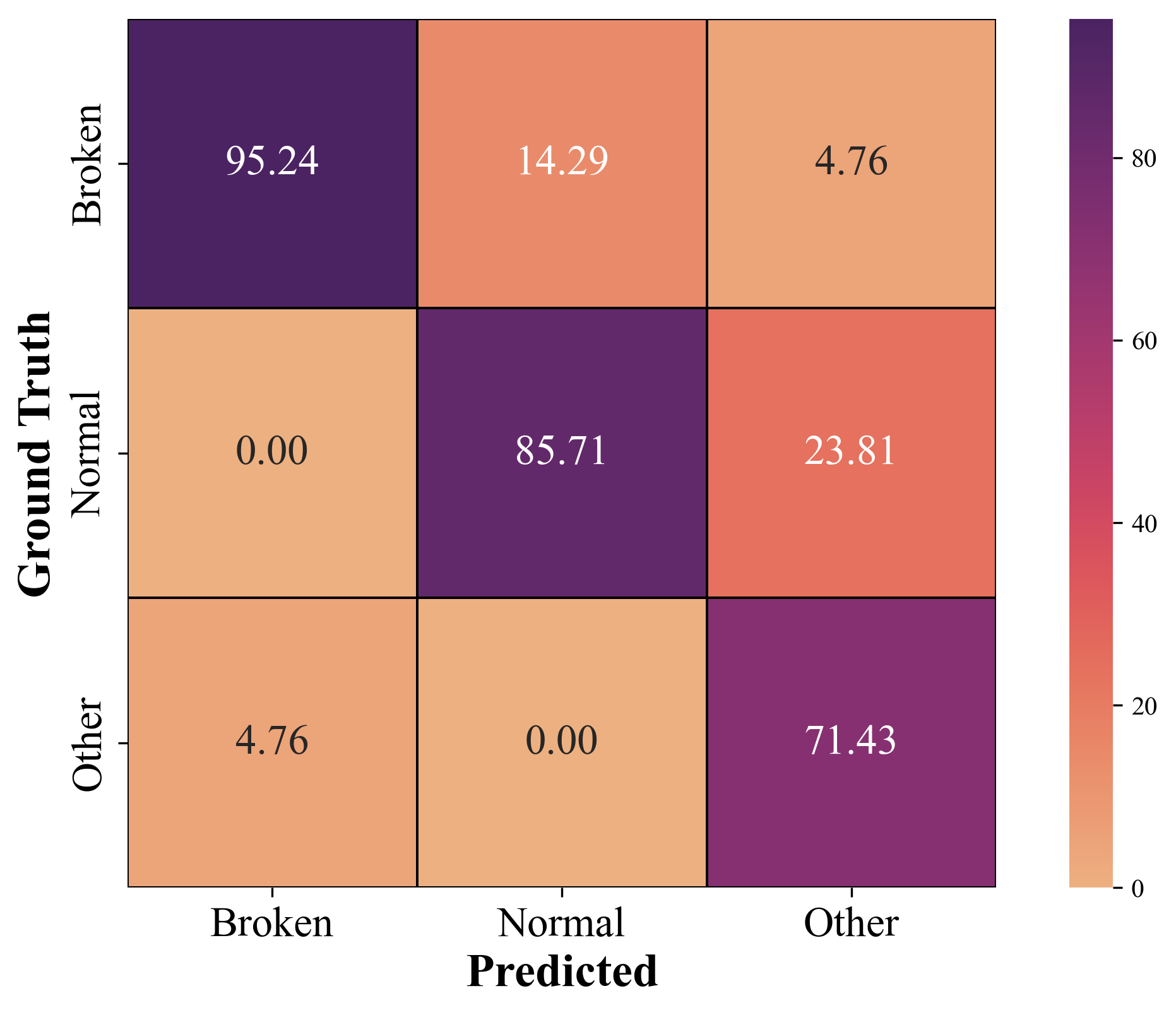}
  \caption{The proposed method (CNN - LSTM - Attention - Leaky ReLU) on the original dataset.}
  \label{fig:result5}
\end{figure}

\begin{table}[]
\centering
\caption{The comparison of the original and augmented datasets using the same proposed method (CNN - LSTM - Attention - Leaky  ReLU).}
\label{tablecompa}

\begin{tabular}{l|c}
\hline
\textbf{Dataset}                  & \textbf{Mean Accuracy (\%)} \\ 
\hline
Augmented dataset (603 sounds)   & 92.35                       \\ \hline
Original dataset (201 sounds)   & 84.13                       \\ \hline
\end{tabular}

\end{table}
\section{Conclusion}
In this article, we proposed a deep learning model for drill fault detection. We applied time-shifting and volume control augmentation methods to increase the number of sounds in the small dataset. The sounds in the augmented dataset were converted into log-Mel spectrograms. These log-Mel spectrograms were used to train our proposed CNN architecture with the Leaky ReLU activation function in conjunction with attention-based BiLSTM for detecting drill failure.
It was found that the overall accuracy of our proposed system reached 92.35\%. This method has a huge potential to be used to diagnose faults in industrial machines. It is a non-invasive method of diagnosing machine failure using short sounds or small datasets. For future work, we consider using both sounds and images to detect the drill failure to obtain better drill fault detection results.

\bibliographystyle{IEEEtran}

\end{document}